\documentclass[aps,floats]{revtex4}
\usepackage{amsmath,amssymb}
\usepackage{graphicx,epsfig}
\usepackage[greek,english]{babel}

\begin{document}
\bibliographystyle {plain}

\def\oppropto{\mathop{\propto}} 
\def\opsimeq{\mathop{\simeq}}
\def\opoverderline{\mathop{\overline}}
\def\operarrow{\mathop{\longrightarrow}}
\def\opsim{\mathop{\sim}}

\def\fig#1#2{\includegraphics[height=#1]{#2}}
\def\figx#1#2{\includegraphics[width=#1]{#2}}


\title{ Star junctions and watermelons of pure or random quantum Ising chains : \\
finite-size properties of the energy gap at criticality } 


\author{ C\'ecile Monthus }
 \affiliation{Institut de Physique Th\'{e}orique, 
Universit\'e Paris Saclay, CNRS, CEA,
91191 Gif-sur-Yvette, France}

\begin{abstract}
We consider $M \geq 2$ pure or random quantum Ising chains of $N$ spins when they are coupled via a single star junction at their origins or when they are coupled via two star junctions at the their two ends leading to the watermelon geometry. The energy gap is studied via a sequential self-dual real-space renormalization procedure that can be explicitly solved in terms of Kesten variables containing the initial couplings and and the initial transverse fields. In the pure case at criticality, the gap is found to decay as a power-law $\Delta_M \propto N^{-z(M)} $ with the dynamical exponent $z(M)=\frac{M}{2}$ for the single star junction (the case $M=2$ corresponds to $z=1$ for a single chain with free boundary conditions) and $z(M)=M-1$ for the watermelon (the case $M=2$ corresponds to $z=1$ for a single chain with periodic boundary conditions). In the random case at criticality, the gap follows the Infinite Disorder Fixed Point scaling $\ln \Delta_M = -N^{\psi} g$ with the same activated exponent $\psi=\frac{1}{2}$ as the single chain corresponding to $M=2$, and where $g$ is an $O(1)$ random positive variable, whose distribution depends upon the number $M$ of chains and upon the geometry (star or watermelon).

\end{abstract}

\maketitle

\section{ Introduction  }

The quantum Ising chain is the basic model for quantum phase transitions at zero temperature 
\cite{sachdev} :
the pure chain is exactly soluble \cite{pfeuty} and is the same universality class as
the two-dimensional classical Ising model via the classical-quantum correspondence \cite{sachdev} ; the random chain is also exactly soluble via the Ma-Dasgupta-Fisher Strong Disorder renormalization approach \cite{fisher_RTFIC} and is the prototype of Infinite Disorder fixed points with unconventional scaling laws with respect to the pure case (see \cite{strongreview} for a review).

Remarkably, for an open Ising chain containing $N$ quantum 
spins with arbitrary transverse fields $h(i) $
and arbitrary couplings $J(i)$
\begin{eqnarray}
H_1^{chain} && = - \sum_{i=1}^N  h(i) \sigma^x(i)
 - \sum_{i=1}^{N-1} J(i) \sigma^z(i) \sigma^z(i+1) 
\label{h1chain}
\end{eqnarray}
there exists an explicit exact formula for the surface magnetization at one end $i=N$
when the spin at the other end $i=1$ is fixed $ $
\cite{peschel,IR1998,dhar,c_micro,strongreview}
\begin{eqnarray}
m^{surf}(N \vert 1) && \equiv <\sigma^z(N) > \vert_{\{\sigma^z(1)=1\}}  
 = \left[ 1+\sum_{i=2}^{N} \prod_{k=i}^N \frac{h^2(k)}{J^{2}(k-1) }   \right]^{-\frac{1}{2}}
\nonumber \\
&& = \left[ 1
+ \frac{h^2(N)}{J^{2}(N-1) } 
+  \frac{h^2(N)}{J^{2}(N-1) }  \frac{h^2(N-1)}{J^{2}(N-2) } 
+...
+  \frac{h^2(N)}{J^{2}(N-1) }  \frac{h^2(N-1)}{J^{2}(N-2) } ... \frac{h^2(2)}{J^{2}(1) }  
 \right]^{-\frac{1}{2}}
\label{msurfzexact}
\end{eqnarray}
For the disordered chain where the transverse fields $h(i)$ and the couplings $J(i)$
are random, 
the surface magnetization thus involves the specific structure of Kesten random variables
  satisfying simple multiplicative recurrence \cite{kes73,kes75,c_micro}.
Kesten random variables also appear either directly
 in many other discrete disordered systems \cite{sol75,sinai,derrida,afa90,hilhorst,calan,us_wettingcayley,us_dynnearzerotrg,oshanin},
or via their continuous analogs given by exponential functionals of Brownian motion
\cite{bou90,flux,yor}.
The probability distribution of the surface magnetization
over disordered samples has thus been analyzed in detail \cite{IR1998,dhar,c_micro}.

Unfortunately, other interesting observables are not given by explicit exact
formulas in terms of the initial disorder variables, but various proposals
have been made to relate them to the known surface magnetizations :

(i) for the chain of Eq. \ref{h1chain} with free boundary conditions, 
the end-to-end correlation 
\begin{eqnarray}
C^{surf}(1,N) && \equiv <\sigma^z(1) \sigma^z(N) > 
\label{corre}
\end{eqnarray}
whose statistics has been computed via the appropriate extension \cite{fisher_gap}
of the Strong Disorder Renormalization \cite{fisher_RTFIC}, was found 
numerically to be very close
to the product of the two surface magnetizations \cite{dhar}
\begin{eqnarray}
C^{surf}(1,N) && \simeq   m^{surf}(N \vert 1) m^{surf}(1 \vert N)
\label{corremsurf}
\end{eqnarray}

(ii) the energy gap between the ground state and the first excited state 
of the chain of Eq. \ref{h1chain} with free boundary conditions
\begin{eqnarray}
\Delta_N \equiv E_1-E_0
\label{gap}
\end{eqnarray}
whose statistics has also been 
studied via Strong Disorder Renormalization \cite{fisher_gap},
was found to be related to the end-to-end correlation of Eq. \ref{corre} via
\begin{eqnarray}
\ln \Delta_N - \ln C^{surf}(1,N) \simeq \sum_{k=1}^{N} \ln h(k)-\sum_{k=1}^{N-1}\ln J(k)
\label{gapcorre}
\end{eqnarray}
as a consequence of the structure of the Strong Disorder renormalization rules  \cite{fisher_gap}.
On the other hand, using the free-fermion approach, the gap was found to be related to the product of the two surface magnetizations
\begin{eqnarray}
 \Delta_N \simeq  m^{surf}(N \vert 1) m^{surf}(1 \vert N)
  \prod_{k=1}^{N-1} \frac{ h(k) }{J(k)}
\label{gapsurf}
\end{eqnarray}
provided $\Delta_N$ decays faster than $1/N$ \cite{aperio}. 
Eq \ref{gapcorre} and  Eq. \ref{gapsurf}
are equivalent if one takes into account Eq. \ref{corremsurf}.

In this paper, we obtain via real-space renormalization some generalizations of these formula for star junctions of $M$ quantum Ising chains, and we use them to analyze the properties of the energy gap.
More precisely, we consider
 $M $ quantum Ising chains containing $N$ spins $\sigma_p(i)$ each
with $1 \leq i \leq N$ and $1 \leq p \leq M$
\begin{eqnarray}
H_M^{chains} && =   \sum_{p=1}^M \left[- \sum_{i=1}^N  h_p(i) \sigma^x_p(i)
 - \sum_{i=1}^{N-1} J_p(i) \sigma_p^z(i) \sigma^z_p(i+1) \right]
\label{hMchains}
\end{eqnarray}
when they are coupled via a
 Star junction involving an additional central spin $\sigma(0)$ 
\begin{eqnarray}
H^{star} && = - h(0)\sigma^x(0)- \sigma^z(0) \sum_{p=1}^M J_p(0) \sigma^z_p(1)+ H_M^{chains}
\label{hstar}
\end{eqnarray}
or when they are coupled  via 
 two Star junctions involving two additional spins $\sigma(0)$ and $\sigma(N+1)$
leading to the Watermelon geometry
\begin{eqnarray}
H^{watermelon} && = - h(0)\sigma^x(0)- \sigma^z(0) \sum_{p=1}^M J_p(0) \sigma^z_p(1)
- h(N+1)\sigma^x(N+1)- \sigma^z(N+1) \sum_{p=1}^M J_p(N) \sigma^z_p(N)
\nonumber \\
&& +H_M^{chains}
\label{hwatermelon}
\end{eqnarray}

The star junction of Eq. \ref{hstar}
with random transverse fields and random couplings has been studied recently via Strong Disorder Renormalization \cite{star_strong} to obtain the disorder-averaged magnetization at the center of the junction $\overline{\sigma^z(0)} $
in the thermodynamic limit of semi-infinite chains $N \to +\infty$.
Pure Quantum Ising chains with uniform transverse field $h_p(i)=h$ and uniform couplings $J_p(i)=J$
when they are coupled via a Delta-junction
\begin{eqnarray}
H^{Delta}&& = -  \sum_{1 \leq p<p' \leq M} J_{p,p'} \sigma^z_{p}(1)\sigma^z_{p'}(1)+ H_M^{chains}
\label{hdelta}
\end{eqnarray}
have also been studied in the limits of weak coupling $J_{p,p'} \to 0$ \cite{star_pure,star_tsvelik} and of strong coupling $J_{p,p'} \to + \infty$ \cite{star_pure}. 
The Delta junction of Eq. \ref{hdelta}
is expected to have the same physical properties as the Star junction of Eq. \ref{hstar}.
 In \cite{star_pure,star_tsvelik}, the case of two Delta junctions 
at the two ends has been also considered and is expected to be similar
to the watermelon geometry of Eq. \ref{hwatermelon} involving instead two star-junctions. Let us finish this introduction by mentioning that other quantum models
have been studied on the star-junction geometry \cite{crampe,star-josephson}.

The paper is organized as follows.
In section \ref{sec_rg}, we introduce a real-space renormalization procedure
for the star junction that can be explicitly solved in terms of 
the surface magnetizations of Eq. \ref{msurfzexact} and other related Kesten variables,
and we obtain generalizations of Eq. \ref{corremsurf} for the correlation between the end spins, and of Eq. \ref{gapsurf} for the energy gap.
In section \ref{sec_pure}, we describe the results for the pure star junction,
and in particular the dynamical exponent $z(M)$ at criticality.
In section \ref{sec_random}, we discuss the results for the disordered star junction
governed by an Infinite Disorder Fixed point.
Section \ref{sec_watermelon} is devoted to the watermelon geometry.
Our conclusions are summarized in section \ref{sec_conclusion}.

\section{ Sequential self-dual real-space renormalization } 

\label{sec_rg}

The Fernandez-Pacheco self-dual real-space
 renormalization introduced initially for the pure quantum Ising chain \cite{pacheco}
has been recently extended to the disordered chain \cite{nishiRandom,c_pacheco,c_renyi}. 
As in any block renormalization procedure, the renormalization rules are applied
{\it in parallel } to all blocks of size $b=2$.
Here we propose to apply the Fernandez-Pacheco self-dual renormalization rules
{\it sequentially } around the central site of the junction of Eq. \ref{hstar}.

\subsection{ Renormalized Hamiltonian }

\begin{figure}[htbp]
 \includegraphics[height=9.4cm]{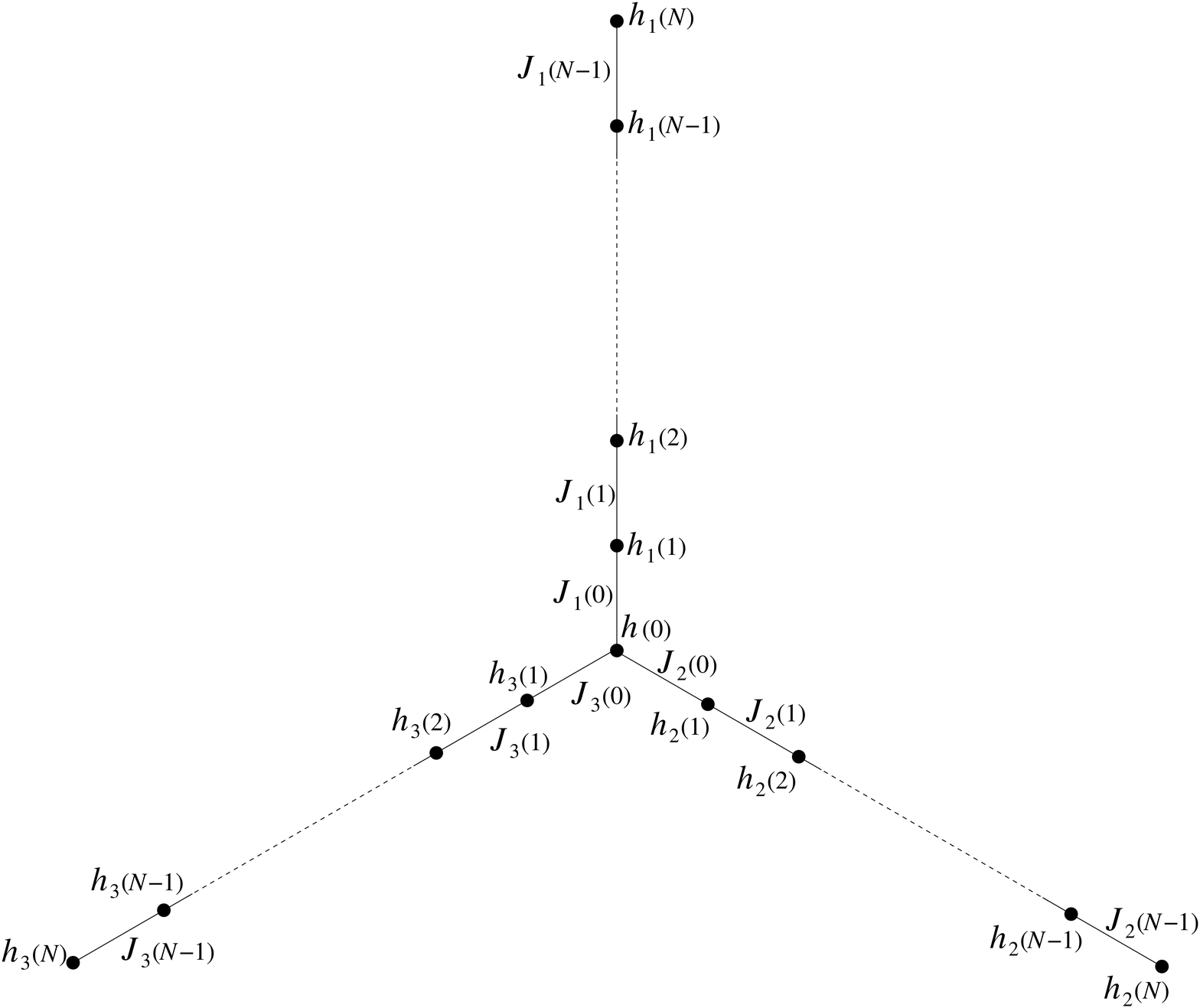}
\hspace{1cm}
 \includegraphics[height=9.4cm]{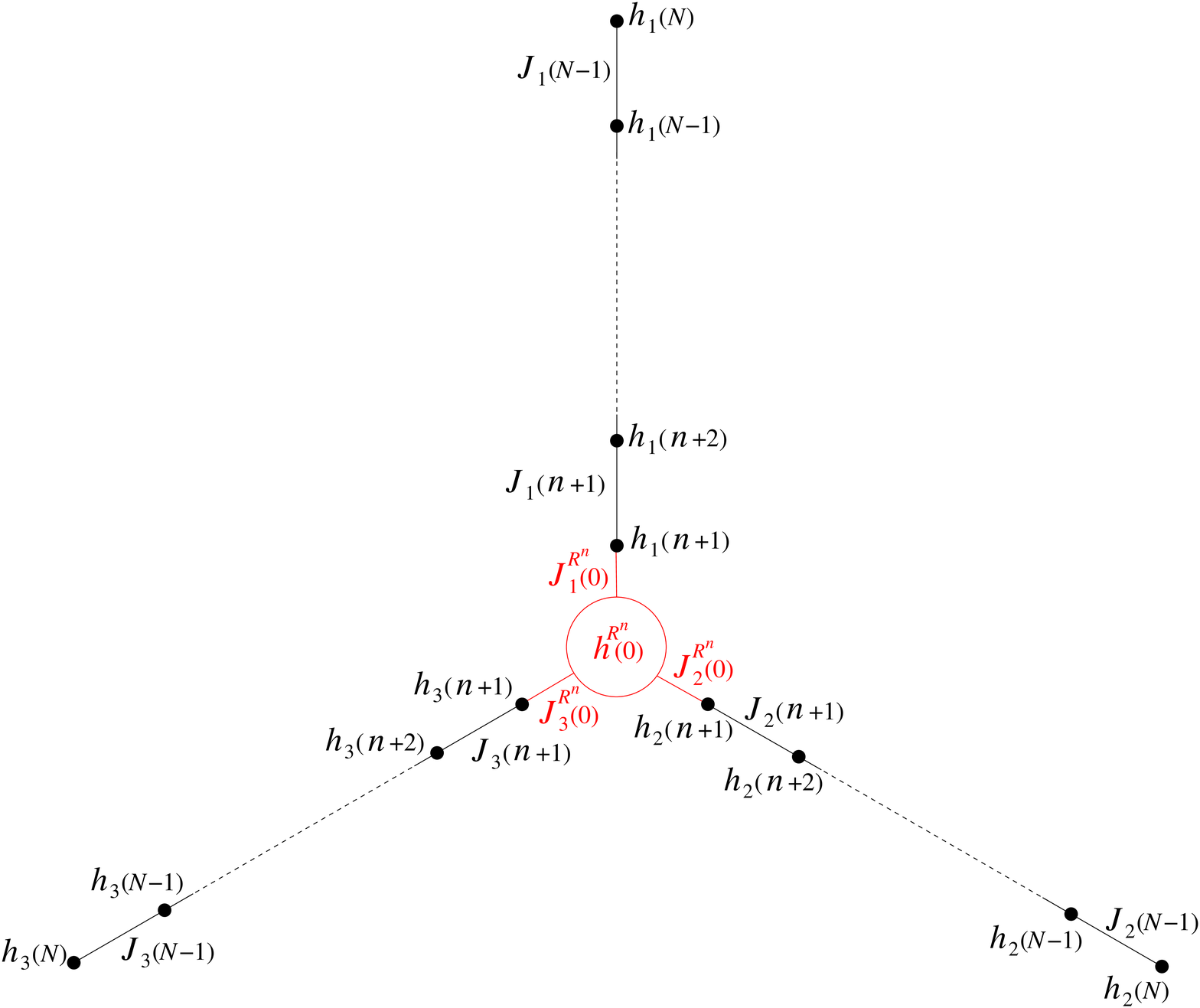}
\caption{ Illustration of the renormalization procedure for the star-junction of $M=3$ chains of length $N$ : \\
Top : Initial Hamiltonian of Eq. \ref{hstar}
with the transverse fields $h_p(i)$ and the couplings $J_p(i)$ \\
Bottom : Renormalized Hamiltonian of Eq. \ref{hstarrg} after $n$ elementary RG steps, where the renormalized center is characterized by the renormalized transverse field $h^{R^n}(0)$ and the renormalized couplings $J_p^{R^n}(0) $. }
\label{figstar}
\end{figure}

More precisely, after $n$ RG steps, 
we consider that the initial Hamiltonian of Eq. \ref{hstar} 
has been renormalized into (see Fig. \ref{figstar})
\begin{eqnarray}
H^{star}_{R^n} && = - h^{R^n}(0)\sigma^x(R^n(0))
- \sum_{p=1}^M J_p^{R^n}(0) \sigma^z(R^n(0))\sigma^z_p(n+1)
\nonumber \\
&& +  \sum_{p=1}^M \left[- \sum_{i=n+1}^N  h_p(i) \sigma^x_p(i)
 - \sum_{i=n+1}^{N-1} J_p(i) \sigma_p^z(i) \sigma^z_p(i+1) \right]
\label{hstarrg}
\end{eqnarray}
where all the spins $n+1 \leq i \leq N$ of the $M$ chains have kept their initial transverse fields $h_p(i) $
and initial couplings $ J_p(i)$ of Eq. \ref{hMchains}, and where the central renormalized spin $\sigma(R^n(0))$
with its renormalized transverse field $h^{R^n}(0)$ and its renormalized couplings $J^{R^n}_p(0)$
replaces the central $(1+Mn)$ initial spins, 
namely the initial central spin and the initial first $n$ spins of the $M$ chains.
The initial Hamiltonian of Eq. \ref{hstar} corresponds to $n=0$ so that the initial condition of the RG flow reads
\begin{eqnarray}
h^{R^{0}}(0) && = h(0)
\nonumber \\
J_p^{R^0}(0)  && = J_p(0)
\label{rgini}
\end{eqnarray}

\subsection{ Renormalization rules }

To go from $ H^{star}_{R^{n-1}}$ to $ H^{star}_{R^{n}} $, we decompose $ H^{star}_{R^{n-1}}$
into the sum of two contributions
\begin{eqnarray}
H^{star}_{R^{n-1}} && = H^{eff} + V
\nonumber \\
H^{eff} && = 
 \sum_{p=1}^M
 \left[ -  J_p^{R^{n-1}}(0) \sigma^z(R^{n-1}(0))\sigma^z_p(n) -   h_p(n) \sigma^x_p(n)
 \right]
\nonumber \\
V && = - h^{R^{n-1}}(0)\sigma^x(R^{n-1}(0))
 +  \sum_{p=1}^M \left[- \sum_{i=n+1}^N  h_p(i) \sigma^x_p(i)
 - \sum_{i=n}^{N-1} J_p(i) \sigma_p^z(i) \sigma^z_p(i+1) \right]
\label{hstarrgsum}
\end{eqnarray}
The Hamiltonian $H^{eff}$ is diagonal in $\sigma^z(R^{n-1}(0)) $,
and for each value $\sigma^z(R^{n-1}(0))=\pm 1 $, one has to diagonalize
independently the single-spin effective Hamiltonian for $\sigma_p(n) $.
We refer to \cite{c_pacheco} for more details and simply state the final output :
the projection of the complementary Hamiltonian $V$
onto the two degenerate ground states of $H^{eff} $
leads to the following RG rules \cite{c_pacheco}
for the central transverse field
\begin{eqnarray}
h^{R^n}(0)  && =h^{R^{n-1}}(0) \prod_{p=1}^M
 \frac{h_p(n)}{\sqrt{h^2_p(n)+(J_p^{R^{n-1}}(0))^2 } }
\label{rgh}
\end{eqnarray}
and for its couplings 
\begin{eqnarray}
J_p^{R^n}(0)  && = J_p(n) \frac{J_p^{R^{n-1}}(0) }{\sqrt{h^2_p(n)+(J_p^{R^{n-1}}(0))^2 } }
\label{rgj}
\end{eqnarray}

\subsection{ Solution for the renormalized couplings }

It is convenient to rewrite Eq \ref{rgj} as
\begin{eqnarray}
\frac{1}{ (J_p^{R^n}(0))^2 }
 && =  \frac{1}{ J_p^2(n) } \left[1+ \frac{h^2_{p}(n)}{(J_p^{R^{n-1}}(0))^2 } \right]
\label{rgjsrb}
\end{eqnarray}
in order to obtain the simple iteration
\begin{eqnarray}
\frac{1}{ (J_p^{R^n}(0))^2 }
 && =  \frac{1}{ J_p^2(n) } \left[1+ \frac{h^2_{p}(n)}{J_p^2(n-1) }
\left( 1+ \frac{h^2_p(n-1)}{(J_p^{R^{n-2}}(0))^2 } \right) \right]=...
\nonumber \\
&& = \frac{1}{ J_p^2(n) } \left[1+ \frac{h^2_{p}(n)}{J_p^2(n-1) }
+ \frac{h^2_{p}(n)}{J_p^2(n-1) }\frac{h^2_{p}(n-1)}{J_p^2(n-2) }+...+
\frac{h^2_{p}(n)}{J_p^2(n-1) }\frac{h^2_{p}(n-1)}{J_p^2(n-2) }...
\frac{h^2_{p}(2)}{J_p^2(1) }\frac{h^2_{p}(1)}{J_p^2(0) } \right]
\nonumber \\
&& =\frac{1}{ J_p^2(n) } \left[1+ 
\sum_{i=1}^n \prod_{k=i}^n  \frac{h_p^2(k)}{J_p^{2}(k-1) }  \right]
\label{jrsol}
\end{eqnarray}
so that the final result reads
\begin{eqnarray}
J_p^{R^n}(0)=  J_p(n)  \left[1+ 
\sum_{i=1}^n \prod_{k=i}^n  \frac{h_p^2(k)}{J_p^{2}(k-1) }  \right]^{-\frac{1}{2}}
\label{jrsolfinal}
\end{eqnarray}
The parenthesis on the right hand side coincides
with the surface magnetization $m^{surf}_{p}(n)$ at $i=n$
of the $p$ chain when the magnetization at the center $i=0$ is fixed
(Eq. \ref{msurfzexact})
\begin{eqnarray}
 m^{surf}_{p}(n \vert 0) && = \left[1+ 
\sum_{i=1}^n \prod_{k=i}^n  \frac{h_p^2(k)}{J_p^{2}(k-1) }  \right]^{-\frac{1}{2}}
\label{mz0p}
\end{eqnarray}
So the solution obtained in Eq. \ref{jrsolfinal} reads
\begin{eqnarray}
J_p^{R^n}(0) && =  J_p(n) m^{surf}_{p}(n \vert 0)
\label{jmz}
\end{eqnarray}
Note that this formula also suggests that 
the sequential RG that we have described is actually somewhat equivalent
after $n$ steps to the large-block self-dual procedure involving a block a size $n$ studied in Ref \cite{igloiSD}.

\subsection{ Solution for the renormalized transverse field }

It is convenient to consider the ratio obtained from Eqs \ref{rgh} and \ref{rgj}
\begin{eqnarray}
\frac{h^{R^n}(0)  }{ \prod_{p=1}^M J_p^{R^n}(0)} && = 
\left[ \prod_{p=1}^M \frac{h_p(n)}{ J_p(n) }  \right]
 \frac{h^{R^{n-1}}(0)  }{  \prod_{p=1}^M J_p^{R^{n-1}}(0)}
\label{rgratiosr}
\end{eqnarray}
in order to obtain the simple iteration
\begin{eqnarray}
\frac{h^{R^n}(0)  }{ \prod_{p=1}^M J_p^{R^n}(0)} && = 
\left[ \prod_{p=1}^M \frac{h_p(n)}{ J_p(n) }  \right]
\left[ \prod_{p=1}^M \frac{h_p(n-1)}{ J_p(n-1) }  \right]
 \frac{h^{R^{n-2}}(0)  }{  \prod_{p=1}^M J_p^{R^{n-2}}(0)} =...
\nonumber \\
&& = \left[ \prod_{p=1}^M \frac{h_p(n)}{ J_p(n) }  \right]
\left[ \prod_{p=1}^M \frac{h_p(n-1)}{ J_p(n-1) }  \right] ...
\left[ \prod_{p=1}^M \frac{h_p(1)}{ J_p(1) }  \right]
 \frac{h(0)  }{  \prod_{p=1}^M J_p(0)}
\label{rghsr}
\end{eqnarray}

The solution obtained in Eq \ref{jmz} for the renormalized couplings
in terms of the surface magnetizations of Eq. \ref{mz0p}
yields for the renormalized transverse field
\begin{eqnarray}
h^{R^n}(0)  = \left[ \prod_{p=1}^M  m^{surf}_{p}(n \vert 0) \right]
\left( \frac{h(0) \prod_{p=1}^M \prod_{i=1}^n h_p(i)}{\prod_{p=1}^M \prod_{i=0}^{n-1} J_p(i) }
\right)
\label{hmz}
\end{eqnarray}

Using the explicit form of the surface magnetization of Eq. \ref{mz0p},
this can be also rewritten as
\begin{eqnarray}
&& h^{R^n}(0)  = h(0) \prod_{p=1}^M \left(
  \frac{h_p(1) h_p(2) ...h_p(n)}{ J_p(0) J_p(1) ... J_p(n-1)}  m^{surf}_{p}(n \vert 0) \right)
\nonumber \\
&& = h(0) \prod_{p=1}^M \left(  \frac{h_p(1) h_p(2) ...h_p(n)}{ J_p(0) J_p(1) ... J_p(n-1)}
  \left[1+ \frac{h^2_{p}(n)}{J_p^2(n-1) }
+ \frac{h^2_{p}(n)}{J_p^2(n-1) }\frac{h^2_{p}(n-1)}{J_p^2(n-2) }+...+
\frac{h^2_{p}(n)}{J_p^2(n-1) }...
\frac{h^2_{p}(1)}{J_p^2(0) }
  \right]^{-\frac{1}{2}} \right)
\nonumber \\
&& =  h(0) \prod_{p=1}^M 
  \left[\frac{ J_p^2(0) J_p^2(1) ... J_p^2(n-1)}{h_p^2(1) h_p^2(2) ...h_p^2(n)}
+ \frac{ J_p^2(0) J_p^2(1) ... J_p^2(n-2)}{h_p^2(1) h_p^2(2) ...h_p^2(n-1)}
+ ...
+ \frac{J_p^2(0) }{h^2_{p}(1)}
+1  \right]^{-\frac{1}{2}} 
\nonumber \\
&& =  h(0) \prod_{p=1}^M 
  \left[1+ \sum_{i=1}^{n} \prod_{k=1}^i  \frac{J_p^2(k-1) }{h^2_{p}(k)}
  \right]^{-\frac{1}{2}} 
\label{rghsrfinal}
\end{eqnarray}

\subsection{ Gap of the star junction with free boundary conditions }

For the finite star junction of Eq. \ref{hstar}, the renormalization procedure described above ends after $N$ RG steps where the renormalized Hamiltonian of Eq. \ref{hstarrg}
contains a single renormalized spin $\sigma(R^N(0)) $ representing the whole star junction
\begin{eqnarray}
H^{star}_{R^N} && = - h^{R^N}(0)\sigma^x(R^N(0))
\label{hstarrgfinal}
\end{eqnarray}
The remaining renormalized transverse field $h^{R^N}(0) $
corresponds to the energy scale associated to the global flip of the junction
and thus represents the energy gap (up to an unimportant factor of $2$)
\begin{eqnarray}
\Delta_N^{star} = h^{R^N}(0)
\label{gapstar}
\end{eqnarray}

Eq. \ref{hmz} yields
\begin{eqnarray}
\Delta_N^{star}  = \left[ \prod_{p=1}^M  m^{surf}_{p}(N \vert 0) \right]
\left( \frac{h(0) \prod_{p=1}^M \prod_{i=1}^N h_p(i)}{\prod_{p=1}^M \prod_{i=0}^{N-1} J_p(i) }
\right)
\label{gapmz}
\end{eqnarray}
This formula is thus the analog of Eq. \ref{gapsurf} :
the numerator contains
the product of the $M$ surface magnetizations and
 the product of all the transverse fields,
whereas the denominator involves the product of all the couplings.

The alternative form of Eq. \ref{rghsrfinal} reads
\begin{eqnarray}
\Delta_N^{star}  = h(0) \prod_{p=1}^M 
  \left[1+ \sum_{i=1}^{N} \prod_{k=1}^i  \frac{J_p^2(k-1) }{h^2_{p}(k)}
  \right]^{-\frac{1}{2}} 
\label{gapexpli}
\end{eqnarray}

\subsection{ Correlation between the end spins }

Before the last Hamiltonian of Eq. \ref{hstarrgfinal},
the renormalized Hamiltonian at step $n=N-1$ (Eq. \ref{hstarrg})
\begin{eqnarray}
H^{star}_{R^{N-1}} && = - h^{R^{N-1}}(0)\sigma^x(R^{N-1}(0))
- \sum_{p=1}^M J_p^{R^{N-1}}(0) \sigma^z(R^{N-1}(0))\sigma^z_p(N)
- \sum_{p=1}^M    h_p(N) \sigma^x_p(N)
\label{hrbeforelast}
\end{eqnarray}
involves the initial end-spins $\sigma_p(N) $ of the $M$ chains,
and the renormalized central spin $\sigma(R^{N-1}(0)) $ 
representing the rest of the sample.
The eigenfunction corresponding to these two last RG steps reads
(see \cite{c_pacheco} for more details on the eigenfunctions of $H^{eff}$ of Eq.
 \ref{hstarrgsum})
\begin{eqnarray}
\psi(S_0; S_p(N)) =  \frac{(\delta_{S_0,1}+\delta_{S_0,-1} )}{\sqrt 2} 
\prod_{p=1}^M \left[ \sqrt{ \frac{1+\frac{J_p^{R^{N-1}}(0) }{  \sqrt{ h_p^2(N)+(J_p^{R^{N-1}}(0))}}}{2} } \delta_{S_p(N),S_0}
+  \sqrt{ \frac{1-\frac{J_p^{R^{N-1}}(0) }{  \sqrt{ h_p^2(N)+(J_p^{R^{N-1}}(0))}}}{2} } \delta_{S_p(N),-S_0}  \right]
\label{psilast}
\end{eqnarray}
The joint distribution of the $M$ end-spins $S_p(N)$
is thus obtained after the integration of the renormalized central spin
\begin{eqnarray}
&& {\cal P}(S_p(N))  = \sum_{S_0=\pm 1} \psi^2(S_0; S_p(N))
\label{endjoint} \\
&& = 
 \sum_{S_0=\pm 1} \frac{1}{ 2} (\delta_{S_0,1}+\delta_{S_0,-1} )
\prod_{p=1}^M \left[ \frac{1+\frac{J_p^{R^{N-1}}(0) }{  \sqrt{ h_p^2(N)+(J_p^{R^{N-1}}(0))}}}{2}  \delta_{S_p(N),S_0}
+   \frac{1-\frac{J_p^{R^{N-1}}(0) }{ \sqrt{  h_p^2(N)+(J_p^{R^{N-1}}(0))}}}{2}  \delta_{S_p(N),-S_0}  \right]
\nonumber 
\end{eqnarray}
In particular, the correlation between the $M$ end-spins reduces to
\begin{eqnarray}
< \prod_{p=1}^M S_p(N) > = 
\prod_{p=1}^M \frac{J_p^{R^{N-1}}(0) }{  \sqrt{ h_p^2(N)+(J_p^{R^{N-1}}(0))}} 
\label{correend}
\end{eqnarray}
and can be thus rewritten using Eqs \ref{mz0p} and \ref{jmz}
as the product of the $M$ surface magnetizations when the center is fixed
\begin{eqnarray}
< \prod_{p=1}^M S_p(N) > = 
\prod_{p=1}^M m^{surf}_p (N \vert 0)
\label{correendsurf}
\end{eqnarray}
This formula is thus somewhat the analog of Eq. \ref{corremsurf}.

\section{ Results for the pure junction of $M \geq 2$ chains }

\label{sec_pure}

For the pure case, the control parameter of a single chain is the ratio between the 
ferromagnetic coupling $J$ and the transverse field $h$
\begin{eqnarray}
K \equiv \frac{J}{h}
\label{Kpur}
\end{eqnarray}
The exact surface magnetization of Eq. \ref{msurfzexact} reads
\begin{eqnarray}
m^{surf}(N \vert 1) && 
= \left[  \frac{1-K^{-2} }{1-K^{-2N} }    \right]^{\frac{1}{2}}
\label{msurfpure}
\end{eqnarray}

\subsection{ Ferromagnetic phase $K>1$ }

In the ferromagnetic phase $K>1$, the surface-magnetization of Eq. \ref{msurfpure}
converges towards the finite limit
\begin{eqnarray}
m^{surf}(N \to +\infty ) && 
= \left[ 1-K^{-2}     \right]^{\frac{1}{2}} \propto_{K \to 1^+} (K-1)^{\beta_s}
\ \ \ { \rm with} \ \ \beta_s=1/2
\label{msurfpureferro}
\end{eqnarray}
As a consequence the renormalized coupling of Eq. \ref{jmz} 
also converges towards a finite limit
\begin{eqnarray}
J_p^{R^N}(0) && = J m^{surf}(N \vert 0) \oppropto_{N \to +\infty}   J \left[ 1-K^{-2}     \right]^{\frac{1}{2}}
\label{jmzpureferro}
\end{eqnarray}
whereas the renormalized transverse field of Eq. \ref{rghsrfinal}
converges exponentially towards zero as
\begin{eqnarray}
h^{R^N}(0)  = h 
  \left[ \frac{K^{2(N+1)}-1}{K^2-1}  \right]^{-\frac{M}{2}} 
\oppropto_{N \to +\infty} K^{- N M}
\label{gapexplipureferro}
\end{eqnarray}

\subsection{ Paramagnetic phase $K<1$ }

In the paramagnetic phase $K<1$, the surface magnetization of Eq. \ref{msurfpure}
decays exponentially with the distance 
\begin{eqnarray}
m^{surf}(N \vert 1) && 
= \left[  \frac{K^{-2}-1 }{K^{-2N}-1 }    \right]^{\frac{1}{2}}
\oppropto_{N \to +\infty} K^N = e^{N \ln (K)}
\label{msurfpurepara}
\end{eqnarray}
 and involves the correlation length exponent $\nu=1$. As a consequence, the renormalized
 coupling of Eq. \ref{jmz} behaves similarly
\begin{eqnarray}
J_p^{R^N}(0) && = J m^{surf}(N \vert 0) \oppropto_{N \to +\infty} K^N
\label{jmzpurepara}
\end{eqnarray}
whereas the renormalized 
transverse field of Eq. \ref{rghsrfinal} converges to the finite limit
\begin{eqnarray}
 h^{R^{N}}(0) =  h  
  \left[ \frac{1-K^{2(N+1)}}{1-K^2}  \right]^{-\frac{M}{2}}  
\oppropto_{N \to +\infty} h(1-K^2)^{\frac{M}{2}}
\label{hmxpara}
\end{eqnarray}

\subsection{ Critical point $K_c=1$  }

At the critical point $K_c=1$, the surface magnetization of Eq. \ref{msurfpure}
decays as the power-law
\begin{eqnarray}
m^{surf}(N \vert 1) && 
= N^{-\frac{1}{2}}
\label{msurfpurecriti}
\end{eqnarray}
As a consequence, the renormalized
 coupling of Eq. \ref{jmz} decays with the same exponent
\begin{eqnarray}
J_p^{R^N}(0) && \propto N^{-\frac{1}{2}}
\label{jmzpurecriti}
\end{eqnarray}
whereas the renormalized 
transverse field of Eq. \ref{rghsrfinal} representing the gap
decays as the power-law
\begin{eqnarray}
\Delta_N =  h^{R^{N}}_{K_c=1}(0) \propto N^{- z(M)}
\label{hmxcriti}
\end{eqnarray}
with the dynamical exponent
\begin{eqnarray}
z( M \geq 2) = \frac{M}{2}
\label{zM}
\end{eqnarray}
For $M=2$ corresponding to a single open chain with free boundary conditions,
this result reduces to $z(M=2)=1$ as it should.
For $M>2$, Eq. \ref{zM} is in agreement with 
 the strong-coupling perturbative expansion of Eq. (12) of Reference \cite{star_pure}.

\section{ Results for the disordered junction of $M \geq 2$ chains}

\label{sec_random}

For the disordered junction, 
the usual control parameter  \cite{fisher_RTFIC}
\begin{eqnarray}
\delta \equiv \frac{\overline{\ln h} - \overline{\ln J}}{var(\ln h) + var(\ln J)}
\label{delta}
\end{eqnarray}
valid for Gaussian distributions can be more precisely defined as the root of the equation
 \cite{igloiexpdelta1,igloiexpdelta2,igloiexpdelta3}
\begin{eqnarray}
\overline{ \left( \frac{J}{h} \right)^{ 2\delta}} =1 
\label{deltapower}
\end{eqnarray}
that makes the link with the literature on Kesten variables
  \cite{kes73,kes75,sol75,sinai,derrida,afa90,hilhorst,calan}.
It is also usual in the field to redefine the length scale to absorb the variance
\cite{fisher_RTFIC}, in order to have the single parameter $\delta$.

\subsection{ Reminder on the statistical properties of the surface magnetization   }

\label{subsec_surf}

Let us briefly summarize the known results on the statistics of the surface magnetization
 of random chains \cite{IR1998,dhar,c_micro,strongreview} in
 the universal critical region around $\delta=0$ :

(i) in the paramagnetic phase $\delta>0$, the surface magnetization reads  \cite{dhar,c_micro}
\begin{eqnarray}
\ln m_{\delta>0}^{surf}(N) = - 2 \delta N + \sqrt{N} v
\label{mspara}
\end{eqnarray}
where $v$ is a zero-mean random variable distributed with the Gaussian distribution
\begin{eqnarray}
G(v) = \frac{1}{\sqrt{4 \pi}} e^{- \frac{v^2}{4} }
\label{vpara}
\end{eqnarray}
In particular, the typical value
\begin{eqnarray}
\overline{ \ln m_{\delta>0}^{surf}(N)} = - 2 \delta N 
\label{avmspara}
\end{eqnarray}
involves the typical correlation length exponent $\nu_{typ}=1$,
whereas the finite-size scaling properties determined by the balance between the two 
terms of Eq. \ref{mspara} are governed by $\nu_{FS}=2$.

(ii) at criticality $\delta=0$, the surface magnetization reads  \cite{dhar,c_micro}
\begin{eqnarray}
\ln m_{\delta=0}^{surf}(N) = -  \sqrt{N} w
\label{mscriti}
\end{eqnarray}
where $w$ is a positive random variable distributed with the half-Gaussian distribution
\begin{eqnarray}
P(w) = \frac{\theta(w \geq 0)}{\sqrt{ \pi}} e^{- \frac{W^2}{4} }
\label{wcriti}
\end{eqnarray}

(iii) in the ferromagnetic phase $\delta<0$, the surface magnetization reads  \cite{dhar,c_micro}
\begin{eqnarray}
\ln m_{\delta<0}^{surf}(N) = -  B
\label{msferro}
\end{eqnarray}
where $B$ is a positive random variable distributed with the exponential distribution
\begin{eqnarray}
Q(B) = \theta(B \geq 0) \mu e^{- \mu B} \ \ \ {\rm  with } \ \ \ \mu \equiv -2 \delta
\label{bferro}
\end{eqnarray}

From these known results concerning the surface magnetization of a single random chain,
 we may now derive the statistical properties of the renormalized
parameters for the random star junction.

\subsection{ Random star junction at the critical point $\delta=0$  }

At the critical point $\delta=0$, 
the renormalized coupling $J^{R^N}$ 
behaves as the surface magnetization of Eq. \ref{mscriti}
\begin{eqnarray}
\ln J^{R^N}_{\delta=0} \propto \ln m_{\delta=0}^{surf}(N) && = -  \sqrt{N} w
\label{jcritirandom}
\end{eqnarray}
The gap $\Delta_M(N)$ given by the renormalized transverse field  will be also of the form
\begin{eqnarray}
\ln \Delta_M(N) = \ln h^{R^N}_{\delta=0} = -  \sqrt{N} g
\label{gapcritirandom}
\end{eqnarray}
and thus involves the same activated exponent
\begin{eqnarray}
\psi = \frac{1}{2}
\label{psi}
\end{eqnarray}
 as the chain corresponding corresponding to the special case $M=2$.
The variable $g=\sum_{p=1}^M w_p$ in Eq. \ref{gapcritirandom}
 is distributed with the convolution of 
$M$ half-Gaussian distributions (Eq. \ref{wcriti})
\begin{eqnarray}
{\cal P}_M(g) \simeq \left[ \prod_{p=1}^M \int_0^{+\infty} d w_p  \frac{2}{\sqrt{\pi}} e^{- w_p^2} \right] \delta\left(g- \sum_{p=1}^M w_p \right)
\label{halfgaussconvol}
\end{eqnarray}
However this expression is expected to be valid only for sufficiently large $g$,
whereas the region of small $g$ requires a more precise analysis,
as shown by the Strong Disorder calculations of \cite{fisher_gap} concerning the case $M=2$ of the chain.

\subsection{  Random star junction in the Ferromagnetic phase $\delta<0$  }

In the ferromagnetic phase $\delta<0 $, the renormalized coupling $J^{R^N}$
remains finite as the surface magnetization of Eq. \ref{msferro}
\begin{eqnarray}
\ln J^{R^N}_{\delta<0} \propto \ln m_{\delta<0}^{surf}(N) = -  B
\label{jmsferro}
\end{eqnarray}
whereas the renormalized transverse field will behave as
\begin{eqnarray}
\ln h^{R^N}_{\delta=0}  = - 2 \delta M N + \sqrt{ M N} v
\label{hrferrorandom}
\end{eqnarray}
where $v$ is a Gaussian random variable distributed with Eq. \ref{vpara}.

\subsection{  Random star junction in the Paramagnetic phase $\delta>0$  }

In the paramagnetic phase $\delta>0$, the renormalized coupling 
inherits from the behavior of Eq. \ref{mspara}
of the surface magnetization 
\begin{eqnarray}
\ln J^{R^N}_{\delta>0} \propto\ln m_{\delta>0}^{surf}(N) = - 2 \delta N + \sqrt{N} v
\label{jmspara}
\end{eqnarray}
whereas the renormalized transverse field remains finite
\begin{eqnarray}
\ln h^{R^N}_{\delta>0}  = - B
\label{hrpararandom}
\end{eqnarray}
where $B=\sum_{p=1}^M B_p$ is distributed with the convolution of $M$ exponential distributions of Eq. \ref{bferro} 
\begin{eqnarray}
Q_M(B) \simeq \left[ \prod_{p=1}^M \int_0^{+\infty} d B_p  \mu e^{- \mu B}  \right]
 \delta\left(B- \sum_{p=1}^M B_p \right) = \frac{ \mu^M B^{M-1} }{(M-1)!} e^{- \mu B}
\label{expconvol}
\end{eqnarray}

\section{ Study of the Watermelon geometry  }

\label{sec_watermelon}

In this section, we use the results of
 the previous sections concerning the single star junction with free boundary conditions
(Eq. \ref{hstar})
to analyze the watermelon geometry of Eq. \ref{hwatermelon} where the $M$ chains
are coupled by two star junctions at the two ends.

\subsection{ Real-space renormalization for the Watermelon geometry  }

We apply the same renormalization rules as in section \ref{sec_rg}
around each of the two junctions, so that after $\frac{N}{2}$ RG steps,
the renormalized Hamiltonian reads
\begin{eqnarray}
H^{watermelon}_{R^{\frac{N}{2}}} && =  - h^{R^{\frac{N}{2}}}(0)\sigma^x(R^{\frac{N}{2}}(0))
 - h^{R^{\frac{N}{2}}}(N+1)\sigma^x(R^{\frac{N}{2}}(N+1)) - J^{R^{\frac{N}{2}}}_W \sigma^z(R^{\frac{N}{2}}(0))\sigma^z(R^{\frac{N}{2}}(N+1))
\label{hwatermelonr}
\end{eqnarray}
The renormalized transverse fields of the two centers of the junctions 
are obtained from Eq. \ref{rghsrfinal}
\begin{eqnarray}
&& h^{R^{\frac{N}{2}}}(0)  = h(0) \prod_{p=1}^M \left(
  \frac{h_p(1) h_p(2) ...h_p({\frac{N}{2}})}{ J_p(0) J_p(1) ... J_p({\frac{N}{2}}-1)}
  m^{surf}_{p}({\frac{N}{2}} \vert 0) \right)
\nonumber \\
&& =  h(0) \prod_{p=1}^M 
  \left[\frac{ J_p^2(0) J_p^2(1) ... J_p^2({\frac{N}{2}}-1)}{h_p^2(1) h_p^2(2) ...h_p^2({\frac{N}{2}})}
+ \frac{ J_p^2(0) J_p^2(1) ... J_p^2({\frac{N}{2}}-2)}{h_p^2(1) h_p^2(2) ...h_p^2({\frac{N}{2}}-1)}
+ ...
+ \frac{J_p^2(0) }{h^2_{p}(1)}
+1  \right]^{-\frac{1}{2}} 
\nonumber \\
&& =  h(0) \prod_{p=1}^M 
  \left[1+ \sum_{i=1}^{{\frac{N}{2}}} \prod_{k=1}^i  \frac{J_p^2(k-1) }{h^2_{p}(k)}
  \right]^{-\frac{1}{2}} 
\label{rgh0w}
\end{eqnarray}
and 
\begin{eqnarray}
&& h^{R^{\frac{N}{2}}}(N+1) = h(N+1) \prod_{p=1}^M \left(
  \frac{h_p(N) h_p(N-1) ...h_p({\frac{N}{2}}+1)}{ J_p(N) J_p(N-1) ... J_p({\frac{N}{2}}+1)}  m^{surf}_{p}({\frac{N}{2}}+1 \vert N+1) \right)
\nonumber \\
&& =  h(N+1) \prod_{p=1}^M 
  \left[\frac{ J_p^2(N) J_p^2(N-1) ... J_p^2({\frac{N}{2}}+1)}
{h_p^2(N) h_p^2(N-) ...h_p^2({\frac{N}{2}}+1)}
+ \frac{ J_p^2(N) J_p^2(N-1) ... J_p^2({\frac{N}{2}}+2)}
{h_p^2(N) h_p^2(N-1) ...h_p^2({\frac{N}{2}}+2)}
+ ...
+ \frac{J_p^2(N) }{h^2_{p}(N)}
+1  \right]^{-\frac{1}{2}} 
\nonumber \\
&& =  h(N+1) \prod_{p=1}^M 
  \left[1+ \sum_{i=\frac{N}{2}+1}^N \prod_{k=i}^N  \frac{J_p^2(k) }{h^2_{p}(k)}
  \right]^{-\frac{1}{2}} 
\label{rgh0np1}
\end{eqnarray}
whereas their coupling is obtained by summing the contribution of the $M$ chains
\begin{eqnarray}
 J_W^{R^{\frac{N}{2}}} && = \sum_{p=1}^M J_p \left( \frac{N}{2}  \right)
 m^{surf}_{p}({\frac{N}{2}} \vert 0)  m^{surf}_{p}({\frac{N}{2}}+1 \vert N+1) 
\label{jwatermelonr}
\end{eqnarray}
with
\begin{eqnarray}
m^{surf}(\frac{N}{2} \vert 0) &&  
 = \left[ 1+\sum_{i=1}^{\frac{N}{2}} \prod_{k=i}^\frac{N}{2} \frac{h_p^2(k)}{J_p^{2}(k-1) }   \right]^{-\frac{1}{2}}
\nonumber \\
&& = \left[ 1
+ \frac{h_p^2(\frac{N}{2})}{J_p^{2}(\frac{N}{2}-1) } 
+  \frac{h_p^2(\frac{N}{2})}{J_p^{2}(\frac{N}{2}-1) }  \frac{h_p^2(\frac{N}{2}-1)}{J_p^{2}(\frac{N}{2}-2) } 
+...
+  \frac{h_p^2(\frac{N}{2})}{J_p^{2}(\frac{N}{2}-1) }  \frac{h_p^2(\frac{N}{2}-1)}{J_p^{2}(\frac{N}{2}-2) } ... \frac{h_p^2(1)}{J_p^{2}(0) }  
 \right]^{-\frac{1}{2}}
\label{msurf1}
\end{eqnarray}
and
\begin{eqnarray}
m^{surf}(\frac{N}{2}+1 \vert N+1) &&  
 = \left[ 1+\sum_{i=\frac{N}{2}+1}^N \prod_{k=\frac{N}{2}+1}^i \frac{h_p^2(k)}{J_p^{2}(k) }   \right]^{-\frac{1}{2}}
\nonumber \\
&& = \left[ 1
+ \frac{h_p^2(\frac{N}{2}+1)}{J_p^{2}(\frac{N}{2}+1) } 
+  \frac{h_p^2(\frac{N}{2}+1)}{J_p^{2}(\frac{N}{2}+1) }  \frac{h_p^2(\frac{N}{2}+2)}{J_p^{2}(\frac{N}{2}+2) } 
+...
+  \frac{h_p^2(\frac{N}{2})+1}{J_p^{2}(\frac{N}{2}+1) }  \frac{h_p^2(\frac{N}{2}+2)}{J_p^{2}(\frac{N}{2}+2) } ... \frac{h_p^2(N)}{J_p^{2}(N) }  
 \right]^{-\frac{1}{2}}
\label{msurf2}
\end{eqnarray}

\subsection{ Solution for the pure watermelon  }

For the pure case, where the surface magnetization is given by Eq. \ref{msurfpure}
in terms of the control parameter $K=J/h$ of Eq. \ref{Kpur},
the renormalized coupling of Eq. \ref{jwatermelonr} reads
\begin{eqnarray}
 J_W^{R^{\frac{N}{2}}} && =  M J \frac{1-K^{-2} }{1-K^{-2-N} }
\label{jwatermelonrpure}
\end{eqnarray}
whereas the transverse fields of Eqs \ref{rgh0w} and \ref{rgh0np1} 
have the same value
\begin{eqnarray}
 h^{R^{\frac{N}{2}}}(0) = h^{R^{\frac{N}{2}}}(N+1)&& = h  \left( K^{- N}
 \frac{1-K^{-2} }{1-K^{-2-N} } \right)^{\frac{M}{2}}
\label{rgh0wpure}
\end{eqnarray}

In the ferromagnetic phase $K>1$, the renormalized coupling of Eq. \ref{jwatermelonrpure}
converges towards the finite limit
\begin{eqnarray}
 J_W^{R^{\frac{N}{2}}} && \opsimeq_{N \to +\infty}  M J (1-K^{-2} )
\label{jwpureferro}
\end{eqnarray}
and the renormalized transverse field of Eq. \ref{rgh0wpure}
converges exponentially towards zero as
\begin{eqnarray}
h^{R^N}(0) 
\opsimeq_{N \to +\infty} h (1-K^{-2})^{\frac{M}{2}}  K^{- N\frac{M}{2}}
\label{hwpureferro}
\end{eqnarray}

In the paramagnetic phase $K<1$, the renormalized coupling of Eq. \ref{jwatermelonrpure}
decays exponentially with the distance 
\begin{eqnarray}
 J_W^{R^{\frac{N}{2}}} && \opsimeq_{N \to +\infty}  M J (1-K^2) K^{N}
\label{jwpurepara}
\end{eqnarray}
and the renormalized 
transverse field of Eq. \ref{rghsrfinal} converges to the finite limit
\begin{eqnarray}
 h^{R^{\frac{N}{2}}}(0) = h^{R^{\frac{N}{2}}}(N+1)&& = h  \left(1-K^2  \right)^{\frac{M}{2}}
\label{hwpurepara}
\end{eqnarray}

At the critical point $K_c=1$, one obtains the following power-laws
for the renormalized coupling of Eq. \ref{jwatermelonrpure}
\begin{eqnarray}
 J_W^{R^{\frac{N}{2}}} && =  M J \frac{1 }{1+\frac{N}{2} }
\label{jwpurecriti}
\end{eqnarray}
and for the renormalized 
transverse field of Eq. \ref{rghsrfinal} 
\begin{eqnarray}
 h^{R^{\frac{N}{2}}}(0) = h^{R^{\frac{N}{2}}}(N+1)&& = h  \left(\frac{1 }{1+\frac{N}{2} }  \right)^{\frac{M}{2}}
\label{hwpurecriti}
\end{eqnarray}
For $M=2$ corresponding to a single chain with periodic boundary conditions,
both involve the same exponent, that represents the dynamical exponent $z(M=2)=1$
as it should.
However for $M>2$, the coupling of Eq. \ref{jwpurecriti} dominates over the transverse fields of Eq. \ref{hwpurecriti}, so that the last decimation corresponds to the projection onto the two ferromagnetic states $(++)$ and $(--)$ defining the final renormalized spin
representing the whole watermelon, and its renormalized transverse field
\begin{eqnarray}
h^{last}_W = \frac{ h^{R^{\frac{N}{2}}}(0)h^{R^{\frac{N}{2}}}(N+1) }{ J_W^{R^{\frac{N}{2}}} }
= \frac{  h }{M }  \left(\frac{1 }{1+\frac{N}{2} }  \right)^{M-1}
\label{hwlast}
\end{eqnarray}
 represents the gap
\begin{eqnarray}
\Delta_N =h^{last}_W  \propto N^{- z_W(M)}
\label{gappurecriti}
\end{eqnarray}
with the dynamical exponent
\begin{eqnarray}
z_W( M \geq 2) = M-1
\label{zwM}
\end{eqnarray}
For $M>2$, the dynamical exponent of Eq. \ref{zwM} is bigger than
the dynamical exponent of a single star junction of Eq. \ref{zM}.

\subsection{ Solution for the disordered watermelon  }

The statistical properties of the renormalized transverse fields 
of Eqs \ref{rgh0w} and \ref{rgh0np1} are the same as in the star junction
case given in Eqs \ref{gapcritirandom}, \ref{hrferrorandom} and \ref{hrpararandom}
with the replacement $N \to \frac{N}{2}$.

The statistical properties of the renormalized coupling of Eq. \ref{jwatermelonr}
 can be obtained from the properties of the surface magnetization
recalled in section \ref{subsec_surf}.
In the paramagnetic phase $\delta>0$, it decays as
\begin{eqnarray}
 \ln J_W^{R^{\frac{N}{2}}}  = - 2 \delta N + \sqrt{ N} v
\label{jwrandompara}
\end{eqnarray}
where $v$ is the minimum value among $M$ variables drawn with the Gaussian distribution
of Eq. \ref{vpara}.
In the ferromagnetic phase $\delta<0$, it remains finite
\begin{eqnarray}
\ln J_W^{R^{\frac{N}{2}}} = -  B
\label{jwrandomferro}
\end{eqnarray}
where $B$ is minimum among $M$ positives variables drawn with the convolution $Q*Q=\mu^2 B e^{- \mu B}$ of the exponential distribution of Eq. \ref{bferro}.
At criticality $\delta=0$, it decays as
\begin{eqnarray}
 \ln J_W^{R^{\frac{N}{2}}} = -  \sqrt{N} w
\label{jwrandomcriti}
\end{eqnarray}
where $w$ is the minimum value
 among $M$ positives variables drawn with the half-Gaussian distribution
of Eq. \ref{wcriti}.
In particular, the gap of the watermelon involves the same activated exponent $\psi=1/2$
and it is only the scaling distribution of the gap that depends upon $M$.

\section{ Conclusion }

\label{sec_conclusion}

For $M \geq 2$ pure or random quantum Ising chains of $N$ spins coupled via a single star junction 
or via two star junctions at the two ends, we have introduced a sequential self-dual real-space renormalization procedure that can be explicitly solved in terms of Kesten variables containing the initial couplings and and the initial transverse fields. We have described the properties of the renormalized couplings and renormalized transverse fields in the ferromagnetic phase, in the paramagnetic phase and at the critical point. In particular, we have discussed the finite-size properties of the energy gap $\Delta_M$ at criticality :

(i) In the pure case, the dynamical exponent $z$ governing the power-law decay of the gap $\Delta_M \propto N^{-z(M)} $ was found to be $z(M)=\frac{M}{2}$ for the single star junction and $z(M)=M-1$ for the watermelon geometry (for the case $M=2$ corresponding to a single chain with open or periodic
boundary conditions, both dynamical exponents reduce to $z(M=2)=1$ as it should).

(ii) In the random case, the gap was found to follow the Infinite Disorder Fixed Point scaling $\ln \Delta_M = -N^{\psi} g$ with the same activated exponent $\psi=\frac{1}{2}$ as the single chain corresponding to $M=2$, so that the number $M$ of chains and the geometry (star or watermelon) only enter in
the probability distribution of the $O(1)$ random positive variable $g$.

As a final remark, let us mention some link with the recent studies concerning the Shannon and R\'enyi entropies of the ground state wavefunction of the pure
quantum Ising chain \cite{jms2009,jms2010,moore,grassberger,atas_short,atas_long,luitz_spectro,jms2014,alcaraz,c_renyi}. The R\'enyi entropy of index $q$ of the quantum chain amounts to study the binding of the 'Ising book' (see Fig. 4 of \cite{jms2010}) where $M=2q$ half-planes of classical two-dimensional Ising models are glued together.
Taking the anisotropic limit in the other direction leads to the pure quantum star junction discussed in section \ref{sec_pure}. Although the present study does not deal with observables that are directly relevant for the R\'enyi entropy, we feel that the expansion of the dynamical exponent $z(M)=\frac{M}{2}=q$
for $q=1+\epsilon$ around the Ising plane $q=1$
\begin{eqnarray}
z(M=2(1+\epsilon))=1+ \epsilon
\label{zexp}
\end{eqnarray}
suggests that the appropriate finite-size scaling variable could be $L^{ \epsilon}
= e^{u}$ with
\begin{eqnarray}
u= \epsilon \ln L = (q-1) \ln L
\label{ufss}
\end{eqnarray}
For the periodic quantum chain, Gr\'egoire Misguich and Jean-Marie St\'ephan
have replotted their numerical data showned on Fig. 5 of their paper \cite{jms2010} in terms of the variable $u=(q-1)\ln L$ and have concluded that this finite size scaling could be also compatible with their data \cite{private}. 
For the open quantum chain, the double logarithm $(\ln L)^2$ found numerically
at criticality for $q=1$ by Jean-Marie St\'ephan \cite{jms2014}
and the singular logarithmic terms in $(cst) q \frac{\ln L}{(q-1)}$ found at criticality for $q \ne 1$ via Conformal Field Theory in Ref \cite{moore}
also points towards the finite size scaling variable of Eq. \ref{ufss}.

\section*{Acknowledgments}

It is a pleasure to thank Gr\'egoire Misguich and Jean-Marie St\'ephan
 for very useful discussions and for replotting their numerical data
\cite{private} in terms of the variable of Eq. \ref{ufss}.

\end{document}